# World's First Monolithic SiGe QKD Transmitter Chip

**Florian Honz(1), Winfried Boxleitner(1), Mariana Ferreira-Ramos(1), Michael Hentschel(1), Philip Walther(2), Hannes Hübel(1), and Bernhard Schrenk(1)**

*(1)AIT Austrian Institute of Technology, Center for Digital Safety & Security, 1210 Vienna, Austria. Author e-mail address: florian.honz@ait.ac.at*
*(2)University of Vienna, Faculty of Physics, Vienna Center for Quantum Science and Technology (VCQ), 1090 Vienna, Austria.*

**Abstract:** We present a single-chip photonic QKD transmitter fabricated on a silicon platform. We achieve secure-key generation over 45.9km of field-deployed fiber and prove its operation along 32 WDM channels, by sourcing light without III-V materials.

## 1. Introduction

Quantum computation poses a threat to the security and long-term secrecy of data transmission currently protected by asymmetric cryptography. This situation renders a shift towards information-theoretic secure approaches like quantum key distribution (QKD) a necessity for communication networks [1]. A major roadblock towards a ubiquitous QKD solution is the technological incompatibility with consumer electronics: Even though chip-scale QKD demonstrations were shown on monolithic InP [2-4], hybrid InP/silicon [3, 5] and InP/polymer [6] platforms, these photonic integrated circuits (PIC) rely on the use of III-V semiconductors to facilitate a functionally complete quantum-optic toolbox that includes light generation. This further necessitates costly co-assembly and hermetic packaging, which stands in stark contrast to the simplicity and cost-efficiency of monolithic silicon microelectronics (Fig. 1a). Recently, light generation on silicon has been investigated as a trailblazer for a monolithic integrated QKD engine on silicon [7].

As such graceful solution to seamlessly introduce QKD in commodity electronics, this work presents the world's first monolithic silicon chip solution for an optical QKD transmitter engine that achieves a secure-key rate of 655 b/s at a QBER of 5.05% for a BB84 DV-QKD protocol over a 45.9 km field-deployed fiber link. In addition, we prove compatibility of the QKD transmitter PIC towards colorless operation on 32 different ITU-T WDM channels.

## 2. Single-Chip Silicon Optical Transmitter Engine for Polarization-Encoded BB84 QKD

QKD builds on the unidirectional transmission of a quantum-optic signal between Alice and Bob. This architectural setup yields an extra degree of freedom for practical QKD integration closer to the edge since the tree-like topology of capillary networks enables their operators to centralize bulky and costly equipment at suitable locations, such as central offices or edge datacenters. For the provenly secure BB84 protocol the involved single-photon detectors, which are considerably contributing to the overall system complexity due to their cooling requirements, are the prime choice for such a centralized element: These detectors can be pooled and shared among multiple users that are equipped with vastly simplified and therefore inexpensive QKD transmitters. Toward this, our all-silicon QKD transmitter PIC, fabricated on IMEC's iSiPP50G silicon-on-insulator platform, is shown in Fig. 1b. It includes a 90-µm long forward-biased lateral SiGe PIN junction as light emitter. 1D and 2D grating couplers (GC) are employed for out-of-plane coupling at the rear facet of the light source and the output of the QKD transmitter, respectively. The source feeds the state preparation circuit, which encodes the light according to a 4-state polarization-encoded BB84 protocol. For this, a first phase modulation stage (PM1) in combination with 2×2 multi-mode interference coupler (MMI) sets the power ratio of the light in the following two branches of a second phase modulation stage (PM2-X, PM2-Y). Since the 2D-GC at the output is realized as a polarizing beam combiner, these branches can be attributed to the TE and TM fields of the light emitted by the QKD transmitter. As the optical phases of the two branches can be independently set, any polarization state on the Poincaré sphere can be constructed.

The VLI characteristics of the SiGe light source are reported in Fig. 2a for operating temperatures between 25 and 80°C. The light output $L_{1D}$ at the rear facet (1D-GC) shows an emission that reaches -66.9 dBm for a forward current of $I_F$ = 25 mA at $V_F$ = 2.2 V. Together with (*i*) the optical loss of ~19.5 dB due to the long electro-optic modulation stages and the 2D-GC of the PIC and (*ii*) the spectral filtering loss for channel definition, this leads to an average photon number of $\mu$ = 0.015 photons/symbol at a symbol rate of $R_{sym}$ = 100 MHz. However, as we will prove, this will not prevent secure-key generation. Initial calibration of the state preparation circuit before QKD operation is accomplished through the integrated thermo-optic phase shifter (TOPS, Fig. 1b) in combination with co-integrated monitor PIN diodes ($PIN_{1,2}$). By injecting reverse light upon factory calibration, the SiGe source can be re-used as detector, yielding a photocurrent $I_{SiGe}$ to balance the optical power between the TE and TM branches to set the optical phase difference (α, Fig. 2b) via the TOPS. Direct modulation of the forward current $I_F$ allows pulse carving and the transmission of decoy states. Figure 2c shows an exemplary histogram for super-Gaussian carving at half of the symbol width, including random decoy pulses. The emission spectrum $L_{2D}$ of the SiGe source is shown in Fig. 2d for acquisition at the 2D-GC output. The FWHM bandwidth of the light emission covers the entire C-band of the DWDM grid specified in ITU-T G.694.1, meaning that the light source can in principle cater for colorless WDM operation where the quantum channel is selected through an external WDM multiplexer or bandpass filter (BPF).

Figure 1b presents the experimental setup for evaluating the key exchange using the monolithic all-silicon QKD transmitter PIC. We implement the polarization-based BB84 DV-QKD protocol by generating the four states **H**, **V**, **R** and **L** (see inset in Fig. 1b) at $R_{sym}$ = 100 MHz via an arbitrary waveform generator (AWG). The limited number of available AWG channels prevented us from implementing pulse carving during the QKD performance evaluation; Nevertheless, this does not limit the validity of our results since carving as a tool to suppress symbol transitions does not impact the state generation. The output of the 2D-GC is coupled to an ITU-T G.652B compatible single-mode fiber (SMF) interfacing at an angle of 10°. Optical filtering is applied to emulate a WDM demultiplexer in the fiber plant, having a flat-top transmission of ΔΛ = 200 GHz centered at 1550.12 nm (ITU-T channel 34) and an insertion loss of 0.8 dB (*β* in Fig. 2d). Due to the average photon number of <0.1 h*ν*/sym, no additional power leveling is required and the signal was transmitted either through (*i*) a variable optical attenuator (VOA) to test the compatible optical budget (OB), (*ii*) SMF spools or (*iii*) a field-installed fiber link in Vienna (Fig. 3c).

Manual polarization control (PC) is adopted at Bob's QKD receiver to align the polarization-encoded QKD signal to the polarization analyzer (PA), which measures the received state in the **H**/**V** and **R**/**L** bases. We employed two InGaAs single-photon avalanche photodiodes (SPAD) with a detection efficiency of 8%, dark count rates of 251 and 308 cts/s and a dead time of 10 µs. The registered detection events are then off-line processed to estimate the raw- and secure-key rates (RKR, SKR) and quantum bit error ratio (QBER) after temporal filtering at 50% of the symbol width.

## 3. QKD Performance: Colorless Operation and Key Exchange over Field-Installed Fiber

We first investigated the QKD performance as a function of the OB through lightpath loss emulation via the VOA. In order to evaluate possible impacts due to the co-integrated SiGe light source (Λ in Fig. 1b), we injected external laser light ($\lambda_C$ in Fig. 1b) at 1550.12 nm through an alternative input to the state preparation circuit. For this baseline benchmark, we achieve a QBER of 6.38% ± 0.27% at a RKR of 51.2 kb/s for an OB of 0 dB (● in Fig. 2e). We can accommodate up to 16.8 dB of link loss before surpassing the QBER limit of 11% for secure-key extraction [8]. This budget is equivalent to an SMF reach of 61 km, considering a conservative fiber loss of 0.277 dB/km, as it applies to our lab fibers. Switchi.ng to the fully monolithic silicon QKD transmitter with integrated SiGe source (▲) introduces no performance penalty apart from the limitation in dynamic OB range due to a reduced launch of $\mu$ = 0.015 h*ν*/sym. The accomplished OB of Γ = 8.7 dB proves the capability of the silicon light source to sufficiently supply the QKD transmitter. We expect that further improvement of the SiGe source emission towards reaching $\mu$ = 0.1 h*ν*/sym at a 1 GHz rate leads to state-of-the-art RKRs in the Mb/s range [2, 5]. As we currently reach a QBER of 2.68% ± 0.18% at a RKR of 86.1 kb/s in the InGaAs SPAD saturation limit (*ε*, Fig. 2e), these high rates should further contribute to an enhanced secure-to-raw key ratio than currently accomplished.

As a next step, we investigated the impact of possible fiber-induced performance penalties (Fig. 3a) for the SiGe-sourced QKD transmitter, achieving secure-key generation for a fiber reach up to 31 km. Comparison of the facilitated fiber loss of Ξ = 8.5 dB with the earlier back-to-back budget Γ indicates no significant penalty since Ξ ≈ Γ.

We further evaluated the channel-dependent QKD performance to demonstrate how the broadband emission of the SiGe source can be harnessed to provide spectral flexibility for the network integration of QKD using the same colorless and low-cost QKD transmitter PIC at every network user. For this, the thin-film BPF emulating the WDM demultiplexer in the fiber network was replaced by a 200-GHz tunable optical filter with an insertion loss of 3.9 dB (*γ* in Fig. 2d) to test QKD operation over 4.3 km from 1520.25 to 1576.20 nm. The performance shows a RKR of 4.2 kb/s/basis at a QBER of 8.81% ± 0.33% (Fig. 3b) for channel 34, yielding the maximum SKR of 591 b/s/basis. We can generate secure keys over a wide spectral window from the upper S-band to the lower L-band, covering a large number of 32 ITU-T channels on a 200-GHz WDM grid. The secure-key estimation for a BB84 protocol employing four detectors yields a SKR of >1 kb/s for eleven of these channels, each of them providing a sufficiently large number of keys to secure a classical channel capacity of up to 2 Tb/s according to the NIST-limit for AES-GCM key-renewal (i.e., one new 256-bit AES key per 64 GB of data). Twenty additional channels provide enough keys to secure classical data links with capacities up to 100 Gb/s, which is sufficiently high for low-cost commodity applications.

Finally, we evaluated the QKD transmitter performance over a field-deployed fiber link in Vienna, having a reach of 45.9 km and a loss of 16.5 dB (Fig. 3c). Due to the high loss we migrated to NbTiN-based superconducting nanowire single-photon detectors (SNSPD) at Bob's receiver, having a detection efficiency of 65%, dark count rates of 75 and 123 cts/s and a dead time of 50 ns. We noticed additional background coupled along the field-installed link, which has been eliminated through an additional BPF at the link end. We achieved a RKR of 1.55 kb/s at a QBER of 5.05% ± 0.29%, yielding a SKR of 655 b/s. This is enough to secure a classical channel capacity of 1.31 Tb/s. Long-term QKD operation over a full hour confirmed the stability of the SiGe-sourced QKD transmitter (Fig. 3d), as indicated by the nearly constant RKR. Even though the QBER increases slowly due to polarization drift along the fiber link, it can be fully recovered through re-alignment of the input polarization at Bob's receiver (*δ*).

## 4. Conclusion

We have demonstrated the world's first monolithic silicon QKD transmitter PIC with polarization-modulated SiGe light emitter and have proven its capability to generate secure keys over a 45.9 km field-deployed fiber link. In addition, we have harnessed its silicon-native emission to facilitate colorless QKD operation for which the wavelength is passively chosen among 32 C-band channels. We expect that further increase in the SiGe light generation efficiency and reduced losses for BB84 state preparation will lead to comparable performance as traditional III-V sourced QKD works. The monolithic design on a silicon platform, which entirely omits III-V materials, promises a future seamless co-integration of the QKD transmitter with commodity and SiGe RF electronics, such as hand-held devices for 6G.

***Acknowledgement:*** *This work was supported by the ERC under the EU's Horizon-Europe research and innovation programme (GA No 101189009), through the QuantERA II Programme (GA No 101017733) and the Austrian FFG (GA No FO999906040).*

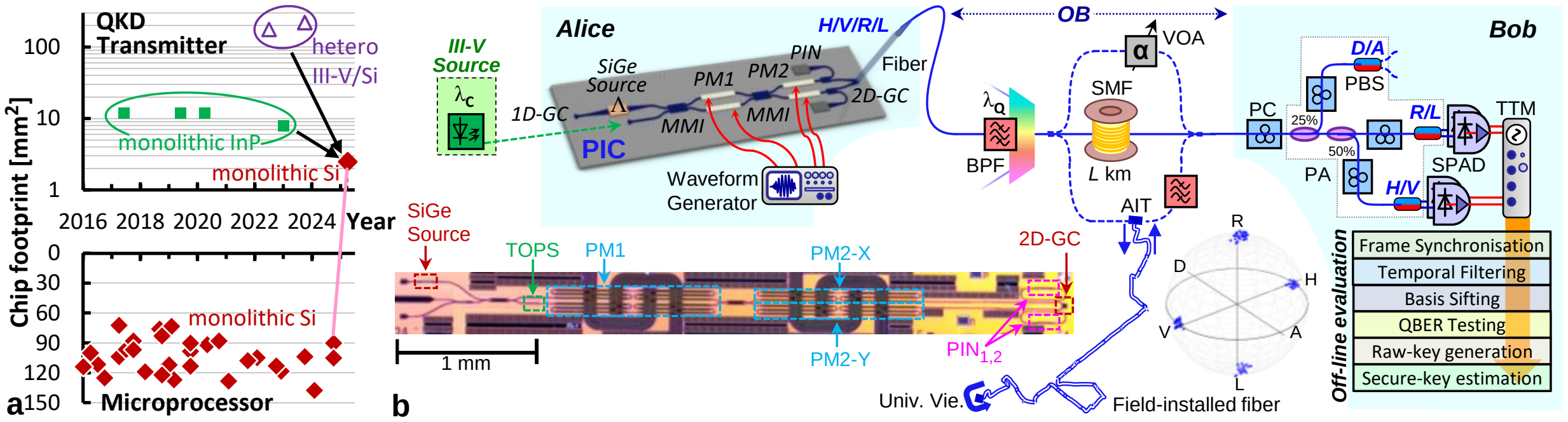

Fig. 1. (a) State of the art in electronic microprocessor and QKD integration. (b) Experimental Setup for all-silicon QKD transmitter evaluation.

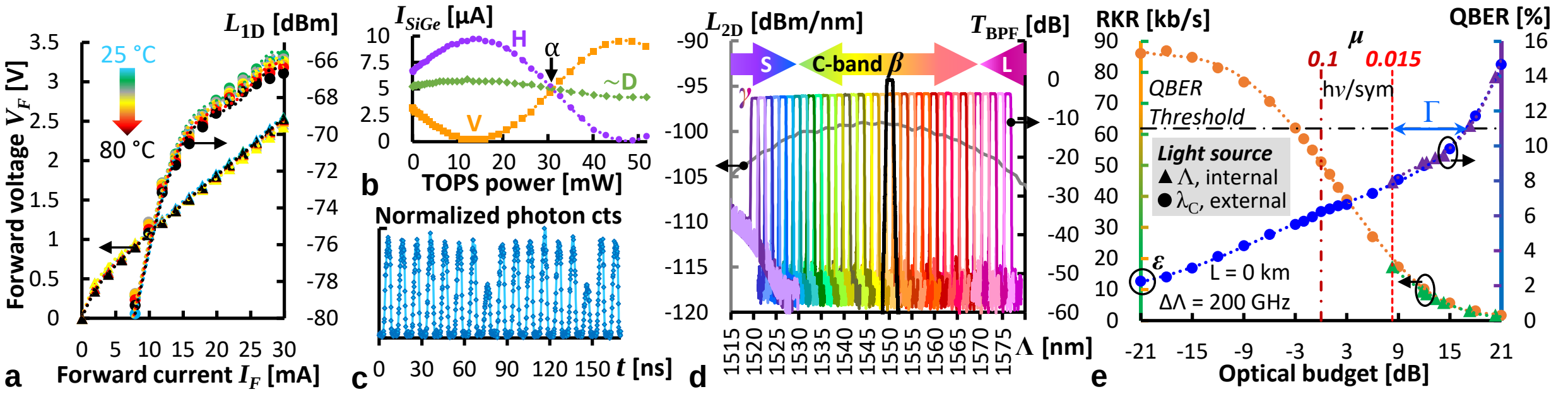

Fig. 2. (a) VLI characteristics of the SiGe source. (b) TOPS calibration via reverse light injection at different polarizations. (c) Source modulation for carving and decoy-state transmission. (d) 2D-GC emission and filter transmission. (e) Compatible OB for externally/internally sourced PIC.

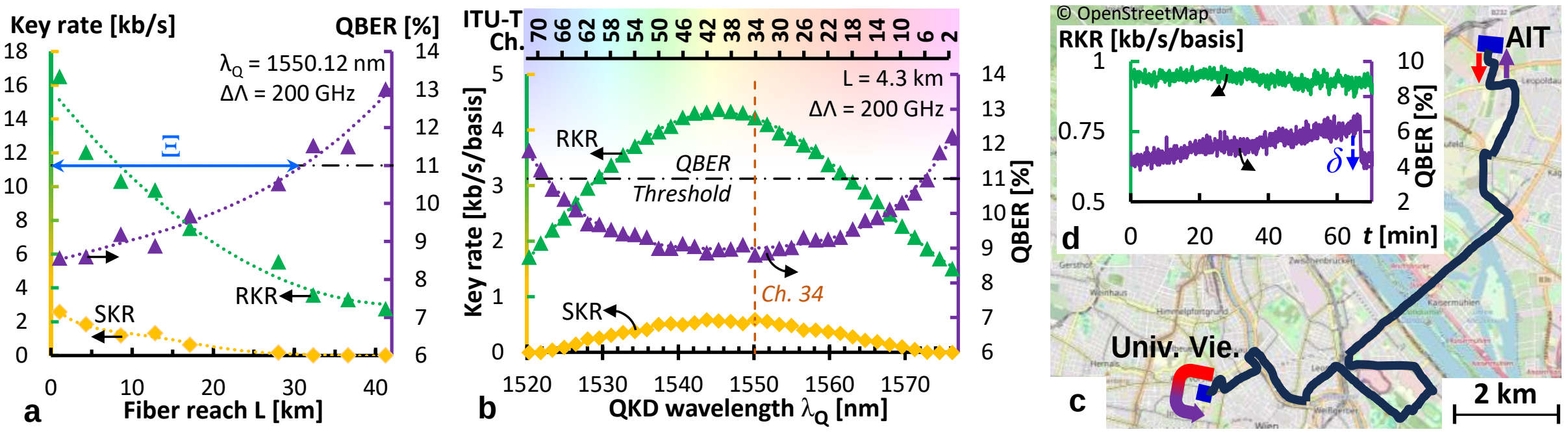

Fig. 3. (a) QKD performance over SMF reach and (b) wavelength channel. (c) Deployed fiber link in Vienna and (d) long-term QKD performance.